\documentstyle[prl,twocolumn,aps]{revtex}
\input{epsf}

\begin{document}
\draft \twocolumn \title{Field quantization for chaotic 
resonators with overlapping modes}

\author{Gregor Hackenbroich, Carlos Viviescas, and Fritz Haake}
\address{Universit\"at Essen, Fachbereich 7, 45117 Essen, 
  Germany}
%\date{ \today}

\maketitle

\begin{abstract}
  Feshbach's projector technique is employed to quantize the
  electromagnetic field in optical resonators with an arbitrary
  number of escape channels. We find spectrally overlapping
  resonator modes coupled due to the damping and noise inflicted by
  the external radiation field. For wave chaotic resonators the mode
  dynamics is determined by a non--Hermitean random matrix. Upon
  including an amplifying medium, our dynamics of open-resonator
  modes may serve as a starting point for a quantum theory of random
  lasing.
\end{abstract}
\vspace*{-0.05 truein} 

Several recent experiments \cite{Cao} have demonstrated laser action
in amplifying random media. The experiments use highly disordered
dielectrics in which light undergoes multiple chaotic scattering.
The scattering can make light stay long enough inside the material
for light amplification to become efficient. A random laser is
created when the amplification rate exceeds the loss rate due to
escape from the medium.

Theoretically, much is known \cite{Bee98} about the sub--threshold
radiation from random lasers but only a few results exist in the
non--linear lasing regime. Progress in this regime has been hampered
by the unusual properties of the random laser modes.  First, the
mode amplitudes and mode frequencies in random lasers depend on the
statistical properties of the underlying random medium. Random laser
modes therefore must be analyzed in a statistical fashion, quite in
contrast to traditional laser resonators. Moreover, the character of
the modes depends on the amount of disorder. For strong disorder,
localization of light may set in and give rise to well separated
modes centered in different regions of space. By contrast, weak
disorder leads to a poor confinement of light and to strongly
overlapping modes.

Standard laser theory \cite{Hak84,Sar74} only applies to
quasi--discrete modes and cannot account for lasing in the presence of
overlapping modes.  Various quantization schemes have been proposed
\cite{Dal99,Lam99,Dut00} to replace the quasi--discrete modes of
standard theory by quasimodes or Fox--Li modes of ``bad'' resonators.
Unfortunately, these schemes are not well suited for a quantum
statistical description required for random lasers.  Statistics
naturally enters the random-scattering theory pioneered by Beenakker
\cite{Bee98}, but that approach is restricted to linear media and
cannot describe lasers above the lasing threshold. So far, to our
knowledge, there is no satisfactory scheme for the field quantization
in random media.

In the present paper we develop such a quantization scheme for
optical resonators with overlapping modes. The resonator may have
an irregular shape or may contain weak random scatterers to ensure
chaotic scattering of light inside the cavity. We employ a
technique previously applied to condensed matter physics, the
Feshbach projector formalism. Using that method we show that the
electromagnetic field Hamiltonian of open resonators reduces to the
well--known system--and--bath Hamiltonian of quantum optics.
Chaotic scattering enters that Hamiltonian in two ways.  First, the
frequency spectrum of the resonator modes shows the correlations
and ``level repulsion'' typical for wave chaos. Second, the
resonator mode amplitudes are not damped separately but coupled by
dissipation. Both effects are related to the spectral properties of
non--Hermitean random matrices and must eventually be included in a
quantum theory of random lasing.

%\iffalse
\begin{figure}[t]
\begin{center}
\leavevmode
\vspace*{-1.1cm}
\epsfxsize = 5.7cm
\epsfbox{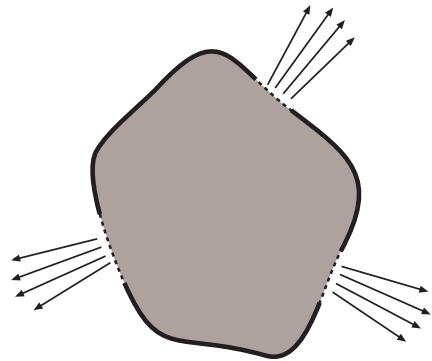}
\end{center}
\vspace*{0.9cm}
\caption{Sketch of a chaotic resonator that is connected to the
  external radiation field via a number of openings.} 
\end{figure}
%\fi

We start with the general solution of the quantization problem, and
then discuss the application to chaotic scattering. For the sake of
simplicity we consider a two dimensional optical resonator and TM
fields; the polarization vector of the electric field defines the
$z$--axis while ${\bf r} =(x,y)$ labels the position in the plane.
The extension to three dimensional resonators and fields with
arbitrary polarization will be given elsewhere \cite{fut1}.
Several openings make for a coupling to the external radiation
field. The total width ${\cal W}$ of the openings determines the
number $M$ of open escape channels at frequency $\omega$, $M \approx
2 {\cal W}/ \lambda$ with $\lambda = 2 \pi c/\omega$. The resonator
modes near frequency $\omega$ are broadened over a frequency range
$M \Delta \omega / 2 \pi$, much greater than their spacing $\Delta
\omega$ if $M \gg 1$. The resonator boundary may have an arbitrary
irregular shape.  For simplicity we assume all walls perfectly
conducting.  The source free Maxwell equations reduce to the scalar
wave equation
\begin{eqnarray}
\left[ \nabla^2 - {1 \over c^2} {\partial^2 \over \partial t^2}  \right]
{A} ({\bf r},t)=0
\label{eq:waveeq}
\end{eqnarray}
for the $z$--component $A({\bf r},t)$ of the vector potential. An
exact quantum description of the total system comprising the
resonator and the external radiation field is obtained by the
so--called modes--of--the--universe approach \cite{Lan73,Gla91}.
One expands both the vector potential ${A}$ and the canonical
momentum field ${\Pi} \equiv \dot{A}/c^2$ in terms of the exact
eigenmodes $\psi_m (\omega,{\bf r})$ of the Helmholtz equation
\begin{eqnarray}
\left[ \nabla^2 + {\omega^2 \over c^2}\right]
\psi_m (\omega,{\bf r})=0 .
\label{eq:emod}
\end{eqnarray}
They are labeled by the continuous frequency $\omega$ and the
integer $m$; the latter specifies the asymptotic conditions far away
from the resonator.  For example, these conditions could correspond
to a scattering problem with incoming and outgoing waves. Then
$\psi_m(\omega,{\bf r})$ represents a solution with an incoming wave
in channel $m$ and only outgoing waves in all other scattering
channels. The channel index $m$ may correspond to an angular
momentum quantum number (for a resonator coupled to free space) or
to a transverse mode index (for resonators connected to external
waveguides).  It is convenient to combine the solutions associated
with the different channels to an $M$--component vector
$\psi(\omega,{\bf r})$. Then the field expansions take the form
\begin{eqnarray}
{A}({\bf r},t) & = & c \int \! d \omega \ q (\omega,t)
\psi(\omega,{\bf r}),
\label{eq:aex1} \nonumber \\
{\Pi}({\bf r},t) & = & {1 \over c} \int \! d \omega \ 
\psi^\dagger (\omega,{\bf r}) p(\omega,t) ,
\label{eq:pex1}
\end{eqnarray}
where the operators $q(\omega)$ and $p(\omega)$ form $M$--component
row and column vectors, respectively. Canonical commutation relation
for $A({\bf r})$ and $\Pi ({\bf r})$ follow by imposing the
canonical commutation relations $[q_m(\omega), p_n(\omega^\prime)]=i
\hbar \delta_{m n}\delta(\omega - \omega^\prime)$. The Hamiltonian
of the problem is given by
\begin{eqnarray}
H = {1 \over 2} \int d {\bf r} \left[ c^2 {\bf \Pi}({\bf r},t)^2
+ [\nabla \times {\bf A}({\bf r},t)]^2 \right],
\label{eq:h}
\end{eqnarray}
with ${\bf A} = A {\bf e}_z$ and ${\bf \Pi} = \Pi {\bf e}_z$. The
Heisenberg equations of motion for ${\bf \Pi}$ and $\nabla \times
{\bf A}$ are easily seen to reduce to the Maxwell equations.

The modes--of--the--universe approach yields a consistent
quantization scheme, but does not provide any information about the
resonator itself. In particular, no definition is obtained for the
resonator modes, and a statistical description of these modes
cannot be implemented at this point.  However, further progress is
possible since the Helmholtz equation with frequency $\omega$ is
equivalent to a single--particle Schr\"odinger equation with energy
$E = \omega^2 / c^2$. That link between optics and single-particle
quantum mechanics allows us to compute the mode functions $\psi
(\omega,{\bf r})$ and to define resonator modes.

The calculation is performed using Fesh\-bach's projector formalism
\cite{Fes62,Dit00}. The single--particle Hilbert space is decomposed
into two orthogonal subspaces associated with the resonator and the
channel region, respectively. The quantum Hamiltonian reads
\begin{eqnarray}
{\cal H} & = & \sum_\lambda E_\lambda | \phi_\lambda \rangle
\langle \phi_\lambda | + \sum_m \int d E E |\chi_m (E)\rangle 
\langle \chi_m (E) |
\nonumber \\& & + \sum_{\lambda m} \int \! \! d E \ \big[ W_{\lambda
m} (E) |\phi_\lambda \rangle \langle \chi_m (E) |
 + {\rm h.c.} \big] \,,
\label{eq:hhd}
\end{eqnarray}
with the first two terms describing the decoupled resonator and
waveguide, while the last term accounts for their coupling. We have
chosen a basis in which both the resonator and the channel
Hamiltonian are diagonal. We note that the Hamiltonian
(\ref{eq:hhd}) is an {\em exact} representation of the eigenvalue
problem (\ref{eq:emod}), even in the regime of overlapping
resonances.

The Hamiltonian (\ref{eq:hhd}) has been extensively used in the theory
of chaotic scattering \cite{Dit00}. We employ the formulation
appropriate for scattering through cavities. The resonator wave
functions $\phi_\lambda({\bf r})$ are nonzero only within the
resonator, while the channel wave functions $\chi (E,{\bf r})$
``live'' only outside. We require that these functions obey Dirichlet
conditions along the boundary of the total system (solid line in
Fig.~1). The boundary condition along the surface separating the
resonator from the waveguide (dashed line in Fig.~1) is arbitrary save
that the total Hamiltonian be self-adjoint.  The coupling amplitudes
$W_{\lambda m}(E)$ are given by surface integrals involving
appropriate resonator and channel wave functions.

Diagonalization of the Hamiltonian (\ref{eq:hhd}) yields the
scattering states $\psi_m(E,{\bf r})$ with incoming wave in channel
$m$ only. The modes functions are found from the mapping to the
Helmholtz equation, $\psi (\omega,{\bf r}) \equiv [\sqrt{2 \omega}
/c] \psi (E,{\bf r})$. They can be expressed as linear combinations
of resonator and channel modes
\begin{eqnarray}
 \psi (\omega,{\bf r}) = \sum_\lambda \alpha_\lambda (\omega)
 \phi_\lambda ({\bf r}) + \int d \omega^\prime \, \beta (\omega,
 \omega^\prime) \chi (\omega^\prime,{\bf r}) ,
\label{eq:psi}
\end{eqnarray}
with an $M$--component coefficient $\alpha_\lambda (\omega)$ and an
$M \times M$ coefficient matrix $\beta(\omega,\omega^\prime)$.
Explicit expressions \cite{Dit00} for these coefficients are not
needed below. Substituting Eq.~(\ref{eq:psi}) into Eqs.\ 
(\ref{eq:pex1}), one obtains the field expansions
\begin{eqnarray}
{A}({\bf r},t) & = & c \sum_\lambda Q_\lambda \phi_\lambda ({\bf
 r}) + c \int \! d \omega^\prime \, Q (\omega^\prime)
 \chi (\omega^\prime ,{\bf r}) ,
\label{eq:aex2} \nonumber \\
{\Pi}({\bf r},t) & = & {1 \over c} \sum_\lambda \phi^*_\lambda ({\bf
  r}) P_\lambda  + {1 \over c} \int \! d \omega^\prime \, 
\chi^\dagger (\omega^\prime ,{\bf r}) P (\omega^\prime)  ,
\label{eq:pex2}
\end{eqnarray}
where we have defined the position operators
\begin{eqnarray}
Q_\lambda  = \! \int \! d \omega \, q(\omega) \alpha_\lambda (\omega) ,
\quad
Q (\omega^\prime)  =  \! \int \! d \omega \, q(\omega)
 \beta (\omega , \omega^\prime),
\label{eq:qoutex}
\end{eqnarray}
and the momentum operators
\begin{eqnarray}
P_\lambda = \! \int \! d \omega \, \alpha^\dagger_\lambda (\omega) 
p (\omega) , \quad
P (\omega^\prime)   = \! \int \! d \omega \,
 \beta^\dagger (\omega , \omega^\prime) p (\omega) .
\label{eq:poutex}
\end{eqnarray}
The final step of the quantization procedure is to express these
operators in terms of photon creation and annihilation operators.
For the resonator modes this is achieved by the representation
\begin{eqnarray}
Q_\lambda & = & \left[{ \hbar \over 2 \omega_\lambda} \right]^{1/2} \, 
\big[ a_\lambda + \sum_{\lambda^\prime} U^\dagger_{\lambda 
\lambda^\prime} 
a^\dagger_{\lambda^\prime}\big] , 
\label{eq:qin} \nonumber \\ 
P_\lambda & = & i
\left[\frac{\hbar \omega_\lambda}{2}\right]^{1/ 2} \, \big[
a^\dagger_\lambda - 
\sum_{\lambda^\prime} U_{\lambda \lambda^\prime} a_{\lambda^\prime}\big] ,
\label{eq:pin}
\end{eqnarray}
where the matrix $U$ with the matrix elements
\begin{eqnarray}
U_{\lambda \lambda^\prime}  =  \int d{\bf r} \phi_\lambda ({\bf r}) 
\phi_{\lambda^\prime} ({\bf r}) 
\label{eq:u1} 
\end{eqnarray}
specifies the spatial overlap of different resonator modes. We note
that $U$ is unitary and symmetric, and that it only couples
degenerate modes, $U_{\lambda \lambda^\prime} \sim
\delta(\omega_\lambda - \omega_{ \lambda^\prime})$, as modes with
different frequencies have zero overlap (the modes are solutions of
an Hermitean eigenvalue problem). The representation (\ref{eq:qin})
realizes the commutation relations and, at the same time, secures
Hermiticity for the intra--cavity fields, $A = A^\dagger$, $\Pi =
\Pi^\dagger$.  A similar representation is obtained for the channel
modes with the replacements $a_\lambda \to b(\omega)$ and
$U_{\lambda \lambda^\prime} \to U(\omega \omega^\prime)$.

Substituting the representation (\ref{eq:pin}) into the field
expansions (\ref{eq:pex2}) and using the unitarity of $U$, we find
the representation of the intra--cavity fields
\begin{eqnarray}
\label{eq:afin}
A({\bf r},t) & = & c \sum_\lambda \left[{ \hbar \over 2
\omega_\lambda}  \right]^{1/2} \, [ a_\lambda \phi_\lambda ({\bf r})
+ a_\lambda^\dagger \phi^*_\lambda ({\bf r}) ] , \nonumber \\
\label{eq:efin}
E({\bf r},t) & = & i \sum_\lambda \left[{ \hbar \omega_\lambda \over
2}  \right]^{1/2} \, [ a_\lambda \phi_\lambda ({\bf r})
- a_\lambda^\dagger \phi^*_\lambda ({\bf r}) ] .
\end{eqnarray}
Substitution of the field expansions into Eq.\ (\ref{eq:h}) finally
yields the field Hamiltonian
\begin{eqnarray}
H & & = \sum_\lambda \hbar \omega_\lambda a^\dagger_\lambda a_\lambda
+ \hbar \sum_m \int \! d \omega \, \omega b^\dagger_m (\omega) 
b_m(\omega) \nonumber \\ 
& & \hspace*{-0.2cm} + \hbar \sum_{\lambda m} \int \! \! d \omega \,
W_{\lambda m} (\omega) \left[ a^{\dagger}_\lambda b_m(\omega) +
a^\dagger_\lambda b_m^\dagger (\omega) + h.c. \right],
\label{eq:hfin}
\end{eqnarray}
where we defined $W_{\lambda m} (\omega) = [c / \hbar \sqrt{ 2
\omega_\lambda}] W_{\lambda m}(E)$ and omitted an irrelevant constant
on the right hand side.  The equations (\ref{eq:afin}) and
(\ref{eq:hfin}) are the key results of the quantization procedure.
The field expansions of the open resonator reduce precisely to the
standard expressions known from closed resonators. However, the field
{\em dynamics} is fundamentally different as shown below. We note that
the resonator modes are coupled to the exter\-nal radiation field via
both resonant ($a^\dagger b$, $ab^\dagger$) and non--resonant ($ab$,
$a^\dagger b^\dagger$) terms. The non-resonant terms can be discarded
here since we are not interested in overdamping (where mode widths
would be larger than or at least comparable to the optical
frequencies).  Our case of interest, the case of overlapping
resonances, is fully compatible with the rotating--wave approximation,
where only the resonant terms are kept.  Then, the Hamiltonian
(\ref{eq:hfin}) reduces to the well--known system--and--bath
Hamiltonian \cite{Gar00} of quantum optics.  It has been argued, that
this Hamiltonian is valid only for good cavities with spectrally
well--separated modes. Our derivation shows that such pessimism is
inappropriate: the system--and--bath Hamiltonian does describe the
dynamics of overlapping modes, provided the broadening of these modes
is much smaller than their frequency (so that non--resonant terms can
be neglected).

We now discuss the field dynamics and address the consequences of
chaotic scattering. As a first example, we establish equations for the
mode amplitudes. From the Hamiltonian (\ref{eq:hfin}) we obtain
\begin{eqnarray}
\label{eq:mot}
\dot{a}_\lambda (t) =   -i \omega_\lambda a_\lambda (t)- \! \pi \!
\sum_{\lambda^\prime} (W W^\dagger)_{\lambda \lambda^\prime}
a_{\lambda^\prime} (t) + F_\lambda (t) ,
\end{eqnarray}
where $F_\lambda (t)$ is the noise operator
\begin{eqnarray}
\label{eq:in} 
F_\lambda (t) = \int_{-\infty}^\infty d \omega 
e^{-i \omega (t -t_0)} \sum_m W_{\lambda m} b_m(\omega,t_0) ,
\end{eqnarray}
and $W$ the coupling matrix with the elements $W_{\lambda m}$. The
equations (\ref{eq:mot}) differ drastically from the
independent--oscillator equations of standard laser theory, in two
respects: First, the mode operators $a_\lambda$ are coupled by the
damping matrix $W W^\dagger$; second, the noise operators
$F_\lambda$ are correlated, $\langle F^\dagger_\lambda
F_{\lambda^\prime} \rangle \neq \delta_{\lambda \lambda^\prime}$, as
different modes couple to the same external channels (the
expectation value is defined with respect to the channel oscillators
at time $t_0$). 

A limiting case of Eq.~(\ref{eq:mot}) is the weak damping regime
where all matrix elements of $W W^\dagger$ are much smaller than the
resonator mode spacing $\Delta \omega$.  This regime can be realized
either by an opening smaller than a wavelength or by the insertion
in the openings of partially reflecting mirrors. To leading order in
$W W^\dagger / \Delta \omega$ only diagonal elements contribute to
the damping matrix, and Eq.~(\ref{eq:mot}) reduces to the standard
equation of motion for non--overlapping modes \cite{Hak84}.

For the interesting case of wave chaos the internal Hamiltonian can be
represented by a random matrix from the Gaussian orthogonal ensemble
of random-matrix theory. The eigenvalues $\omega_\lambda$ display
level repulsion and universal statistical properties. From Eq.\
(\ref{eq:mot}), the mode dynamics of open chaotic resonators is
governed by a {\em non--Hermitean} random matrix; we thus encounter an
interesting connection between the spectral properties of open chaotic
optical resonators and non--Hermitean random matrices \cite{Fyo97}.

As a second application of our field dynamics we now study an open
resonator above the laser threshold. We focus on a single laser--line
and compute the laser linewidth. The active medium is represented by
${\cal N}$ two--level atoms. To simplify the calculation, we assume
(i) ${\cal N}\gg 1$ and spatially uniform gain , (ii) atomic decay
rates that much exceed the field decay rates, (iii) exact resonance
between the laser frequency $\bar{\omega}$ and the atomic transition
frequency, and (iv) laser operation sufficiently far above threshold so
that the field fluctuations can be obtained by linearization. The
Heisenberg equations for the field mode amplitudes and the atomic
polarization and inversion take the standard form \cite{Hak84} {\em
except} for the mode coupling inflicted by damping and noise. We
decompose the field into its classical steady--state value and the
quantum fluctuations, $a_\lambda = (\bar{a}_\lambda + \delta
a_\lambda) \exp( -i \bar{\omega} t)$. The steady--state conditions
take the form $0 = {\cal H} \cdot \bar{a}$ where $\bar{a}$ is an
$N$--component vector comprising the steady--state amplitudes
$\bar{a}_\lambda$ (the limit $N \to \infty$ is taken at the end of the
calculation). The non--Hermitean matrix
\begin{eqnarray}
\label{eq:steady}
{\cal H}_{\lambda \lambda^\prime} = ( \omega_\lambda - \bar{\omega} )
\delta_{\lambda \lambda^\prime} - \pi (W W^\dagger)_{\lambda 
\lambda^\prime} +G \delta_{\lambda \lambda^\prime} ,
\end{eqnarray}
depends on the laser field intensity $I = \sum_\lambda
|\bar{a}_\lambda|^2$ via the gain $G = ( 2 S {\cal N} g^2 /
\gamma_\perp) ( 1 + 4 g^2 I / \gamma_\parallel
\gamma_\perp)^{-1}$. Here, $S$ denotes the pumping strength, $g$ is
the atom--field coupling, and $\gamma_\perp$ and $\gamma_\parallel$
are the decay rates for the atomic polarization and population
inversion. The equations of motion for the quantum fluctuations follow
upon linearization around the steady--state solution
\begin{equation}
\label{eq:fluc}
\left( \begin{array}{c} \delta \dot{a} \\ \delta \dot{a}^\dagger
       \end{array} \right) = {\cal L} \left( \begin{array}{c} \delta a
       \\ \delta a^\dagger \end{array} \right) + \left(
       \begin{array}{c} {\cal F} \\ {\cal F}^\dagger \end{array}
       \right).
\end{equation}
The noise operators ${\cal F}$, ${\cal F}^\dagger$ incorporate both
field noise and noise from the atomic reservoirs. The dynamics of
$\delta a$ and $\delta a^\dagger$ is coupled by the $2N \times 2N$
matrix 
\begin{equation}
\label{eq:linear}
{\cal L}
= 
\left( \begin{array}{cc}
       -i {\cal H}  &  0 \\
       0            &  i {\cal H}^*
       \end{array} \right)
+ \frac {\partial G}{\partial I}
\left( \begin{array}{cc}
  \bar{a} \cdot \bar{a}^\dagger  &  \bar{a} \cdot \bar{a}^T \\
  \bar{a}^* \cdot \bar{a}^\dagger & \bar{a}^* \cdot \bar{a}^T
    \end{array} \right)
\end{equation}
which depends explicitly on the steady--state solution $\bar{a}$.
Equations (\ref{eq:fluc}), (\ref{eq:linear}) reduce the computation of
the field fluctuations to the spectral decomposition of the
non--Hermitean matrix ${\cal L}$. One easily shows that ${\cal L}$ has
a zero eigenvalue connected with the well-known process of phase
diffusion.  The corresponding right eigenvector has the form
$(r,-r^*)$, where $r \propto \bar{a}$ is a right eigenvector to
eigenvalue $0$ of the $N \times N$ matrix ${\cal H}$; the existence of
$r$ and the corresponding left eigenvector $l$ is guaranteed by the
steady--state equations, $0 = {\cal H} \cdot \bar{a}$. The
phase--diffusion coefficient and the laser linewidth $\delta \omega$
can now be computed along standard lines \cite{Hak84,Sar74}: We solve
the equations of motion (\ref{eq:fluc}) and calculate the Fourier
transform of the stationary correlator $\langle \delta
a^\dagger(t)\delta a(0) \rangle$. Keeping only the zero--eigenvalue
contribution in the spectral decomposition of ${\cal L}$, we obtain
the linewidth
\begin{equation}
\label{linewidth}
\delta \omega = K \delta \omega_{\rm ST},
\end{equation}
which is larger than the fundamental (Schawlow--Townes) linewidth
$\delta \omega_{\rm ST}$ by the Petermann factor \cite{Pet79,Sie89}
\begin{equation}
\label{petermann}
K = \langle l | l \rangle \langle r | r \rangle .
\end{equation}
The non--zero eigenvalues of ${\cal L}$ will generally modify the
Lorentzian spectrum and the laser lineshape \cite{Dut99}.  A detailed
investigation of these modifications, their statistics in a random
medium, the pertinent photon statistics \cite{Hac01}, as well as the
generalization to multi--mode lasing will be published separately.

Support by the Sonderforschungsbereich ``Unordnung und gro\ss e
Fluktuationen'' der Deutschen Forschungsgemeinschaft is gratefully
acknowledged.

\end{document}